# Disentangling the magnetoelectric and thermoelectric transport in topological insulator thin films

Jinsong Zhang[1], Xiao Feng[1,2], Yong Xu[3,4], Minghua Guo[1], Zuocheng Zhang[1], Yunbo Ou[2], Yang Feng[1], Kang Li[2], Haijun Zhang[3], Lili Wang[1,5], Xi Chen[1,5], Zhongxue Gan[4], Shou-Cheng Zhang[3,4], Ke He[1,5†], Xucun Ma[1,5], Qi-Kun Xue[1,5], Yayu Wang[1,5†]

[1]*State Key Laboratory of Low Dimensional Quantum Physics, Department of Physics, Tsinghua University, Beijing 100084, P. R. China*

[2]*Institute of Physics, Chinese Academy of Sciences, Beijing 100190, P. R. China*

[3]*Department of Physics, Stanford University, Stanford, CA 94305–4045, USA*

[4]*ENN Intelligent Energy Group, ENN Science Park, Langfang, Hebei 065001, China*

[5]*Collaborative Innovation Center of Quantum Matter, Beijing 100084, China*

[†]Emails: kehe@tsinghua.edu.cn; yayuwang@tsinghua.edu.cn

We report transport studies on $(Bi_{1-x}Sb_x)_2Te_3$ topological insulator thin films with tunable electronic band structure. We find a doping and temperature regime in which the Hall coefficient is negative indicative of electron-type carriers, whereas the Seebeck coefficient is positive indicative of hole-type carriers. This sign anomaly is due to the distinct transport behaviors of the bulk and surface states: the surface Dirac fermions dominate magnetoelectric transport while the thermoelectric effect is mainly determined by the bulk states. These findings may inspire new ideas for designing topological insulator-based high efficiency thermoelectric devices.

## I. INTRODUCTION

Thermoelectric power generation and refrigeration utilize the conversion between heat and electricity carried by electrons in solids, and thus have the advantages of being quiet, reliable, flexible, and eco-friendly [1-4]. The main hurdle for the widespread application of thermoelectricity is its low efficiency compared to conventional technologies. The performance of thermoelectric devices is evaluated by a dimensionless figure of merit defined as $ZT = S^2\sigma T/\kappa$, where $S$, $\sigma$, $T$, and $\kappa$ are the Seebeck coefficient, electrical conductivity, absolute temperature, and thermal conductivity, respectively. Despite extensive efforts in the past few decades, the maximum room-temperature $ZT$ for commercial thermoelectric materials has remained at ~ 1 [3-6]. In recent years, many novel ideas for improving the thermoelectric efficiency have been proposed [3]. A particularly attractive proposal is to improve $ZT$ through nanostructuring, which modifies the electronic structure by quantum size effect and reduces the thermal conductivity by boundary scatterings [2,7]. This idea has inspired continuing experimental efforts to search for the proof-of-principle devices consisting of nanocomposites [6,8], nanowires [9] and superlattices [10,11].

Very recently, $Bi_2Te_3$ and $Sb_2Te_3$, known for decades as the best room-temperature thermoelectric materials, were discovered to be topological insulators (TIs) [12-14]. The topological surface states (SSs) enclosing the TIs provide new opportunity for enhancing their thermoelectric efficiency, because the additional surface transport channel may help circumvent the problems of a single channel where various transport parameters are all entangled and thus cannot be manipulated separately. There have been numerous theoretical discussions on the effect of topological SSs on the thermoelectric behavior of

TIs [15-22]. It has been argued that the topological protection of the SSs leads to superb electrical conductivity [16], thus may give a further boost to *ZT*. On the downside, the Dirac-like linear dispersion of the SSs is not optimal for the Seebeck coefficient, which prefers a sharp change of electron density of state (DOS) near the Fermi level ($E_F$) [23]. There are further complications and meanwhile opportunities caused by the interplay between the surface and bulk states [17], as well as the top and bottom SSs themselves [16,24]. In particular, a recent theoretical work shows that topological SSs can dramatically enhance *ZT* when the $E_F$ is tuned close to the bottom of the bulk conduction band or top of the bulk valence band [17]. In this case, the life-times for quasi-particle excitations above and below the $E_F$ are strongly asymmetrical, leading to opposite sign for the Seebeck and Hall coefficients.

At the current stage, most of these issues have not been investigated adequately by experiments. In this paper we report transport studies on molecular beam epitaxy (MBE)-grown $(Bi_{1-x}Sb_x)_2Te_3$ TI thin films with tunable band structure by varying the Bi/Sb ratio [25]. We find a doping and temperature regime in which the Hall coefficient is negative indicative of electron-type carriers, whereas the Seebeck coefficient is positive indicative of hole-type carriers. This sign anomaly is due to distinct transport behaviors of the bulk and surface states: the surface Dirac fermions dominate magnetoelectric transport while the thermoelectric effect is mainly determined by the bulk states. These findings may inspire new ideas designing TI-based high efficiency thermoelectric devices.

## II. METHODOLOGY

### A. MBE sample growth

The $(Bi_{1-x}Sb_x)_2Te_3$ thin films are grown on insulating $SrTiO_3$ substrates (2 mm × 8 mm × 0.25 mm) by using the same growth method described in our previous report [25]. All the samples have the same thickness $d = 5$ quintuple layers (QLs). Before sample growth, the substrates are degassed at 550 ℃ for 10 min and then heated at 650 ℃ for 25 min in the ultrahigh vacuum chamber. To reduce Te vacancies, the growth is kept in Te-rich condition with the substrate temperature at 200 ℃. Finally, an amorphous Te layer (~10 nm) is deposited on the top to prevent unintentional contamination from ambience.

### B. Transport measurements

Figure 1(a) shows the schematic setup of the transport measurements on the TI films. Electrical transport properties including resistance and the Hall effect are measured in an isothermal condition. Thermoelectric measurements are carried out in high-vacuum condition with the pressure lower than $1 \times 10^{-6}$ mbar. A thin-film heater is mounted on the right end of the substrate to produce the temperature gradient. A pair of fine-gauge thermocouples (type E, CHROMEGA®/Constantan) are connected in subtractive series and thermally anchored to the substrate to monitor the temperature difference. The DC voltages of the Seebeck effect (or the thermopower) and thermocouples are recorded by nano-voltmeters (Keithley 2002 multimeter with 1801 preamplifier), and the Seebeck coefficient of the gold leads for thermoelectric measurement is subtracted.

### C. Theoretical calculations

To carry out *ab initio* calculations of the TI band structure, the BSTATE package [26] with the plane-wave pseudo-potential is employed in the framework of the Perdew–Burke–Ernzerhof-type exchange-correlation functional [27] with generalized gradient approximation of the density functional theory. The kinetic energy cutoff is fixed to 340 eV, and a k-mesh is taken as 8×8×1 for the 5 QL free-standing slab with a 25 Bohr vacuum layer. Virtual crystal method [28] is used to simulate the mixing of Bi and Sb. The lattice constant and the atomic position are obtained by linearly interpolating between $Bi_2Te_3$ and $Sb_2Te_3$. The Seebeck coefficient as a function of temperature and Fermi level is calculated by the Landauer transport approach with the constant mean free path model [29] based on the band structure in a dense k-mesh 400×400×1 from the maximally localized Wannier functions [30].

## III. EXPERIMENTAL RESULTS

We start from pure $Bi_2Te_3$ ($x = 0$) and $Sb_2Te_3$ ($x = 1$) TIs. Fig. 1(b) displays the two-dimensional (2D) sheet resistance ($R_□$) of the two samples, both exhibiting metallic behavior at high $T$ and becoming weakly insulating at $T < 10$ K. These are typical behaviors of TI films with large bulk carrier density and relatively small surface contribution to transport [25]. The temperature dependent $S$ of these two samples are shown in Fig. 1(c), which display a quasi-linear relationship with increasing $T$, as expected for simple metals. The negative $S$ of $Bi_2Te_3$ is because its $E_F$ lies in the bulk conduction band due to the *n*-type bulk carriers induced by Te vacancies, as confirmed by the Hall effect [Fig. 1(d)] and previous angle-resolved photoemission spectroscopy (ARPES) measurements [25]. On the other hand, in $Sb_2Te_3$ the $E_F$ is lower than the top of bulk valence band due to the *p*-type bulk carriers induced by antisite defects, thus leading to a positive $S$ and Hall coefficient $R_H$ [Fig. 1(d)]. At room temperature ($T = 300$ K), $S = $ 120 μV/K for $Sb_2Te_3$ is comparable to the best bulk value for *p*-type thermoelectric [5].

While for $Bi_2Te_3$ the amplitude of $S$ is only about 30 µV/K, much lower than that for the best *n*-type bulk thermoelectric [5]. This can be understood based on the Drude model for free electron gas, in which the Seebeck coefficient is expressed as

$$S = \frac{\pi^2}{6}\left(\frac{k_B T}{E_F}\right)\frac{k_B}{e}. \tag{1}$$

A large carrier density, hence a large $E_F$, could significantly reduce $S$. Indeed, for our $Bi_2Te_3$ film the nominal carrier density estimated from the Hall coefficient is $n_{3D} = 1/eR_H = 1.2 \times 10^{20}$ cm$^{-3}$ at 300 K, which is much larger than the optimal carrier density ($\sim 2 \times 10^{19}$ cm$^{-3}$) for bulk thermoelectric [5].

The transport properties become more interesting as we mix $Bi_2Te_3$ and $Sb_2Te_3$ to form the $(Bi_{1-x}Sb_x)_2Te_3$ ternary compounds. Fig. 2 displays the evolution of $R_\square$ with $T$ for five $(Bi_{1-x}Sb_x)_2Te_3$ films with $0.7 \leq x \leq 0.97$. The $x = 0.95$ sample shows an insulating behavior at low $T$ because in this sample the $E_F$ lies closest to the Dirac point. The resistance behavior becomes less insulating on both sides due to the increase of carrier density. In this regime the Dirac-like SSs make more significant contribution to the electrical transport as reported previously [25].

Figure 3 displays the Hall effect measurements of these five $(Bi_{1-x}Sb_x)_2Te_3$ samples in the temperature range from 1.5 K to 300 K. All the Hall resistance $R_{yx}$ vs. $H$ traces in Fig. 3(a) show linear dependence for magnetic field up to 2 T, and the slope can be used to calculate the Hall coefficient $R_H$ [black squares in Fig. 3(b)]. Generally, when two types of charge carriers coexist in the same system, $R_{yx}$ will become non-linear with magnetic field. The absence of non-linearity indicates the dominance of one type of charge carriers in the Hall effect, which we attribute to the high-mobility surface states

(see below for more detailed discussion). A sign change of the Hall coefficient is clearly observed as Sb concentration $x$ is increased from 0.9 to 0.95. The high-density $R_H$ vs. $T$ data (red lines) displayed in Fig. 1(d) and Fig. 3(b) are taken by the channel switching method at fixed magnetic field [31], and they are in good agreement with that obtained by the linear fit of Hall traces (black squares).

Figure 4 presents the Seebeck coefficients of these five $(Bi_{1-x}Sb_x)_2Te_3$ films. For $x = 0.97$ and 0.95, the Seebeck coefficient behaves similarly to that of pure $Sb_2Te_3$. As reported previously, in these two samples the $E_F$ lies below the Dirac point and close to the top of the bulk valence band [25]. The hole-type carrier density decreases with decreasing $x$, and the Seebeck coefficient becomes larger and reach a high value of 180 μV/K for $x = 0.95$ at $T = 300$ K. This can also be roughly explained by Eq. (1) as a consequence of reduced hole density. As $x$ is reduced to $x = 0.90$, the Seebeck coefficient remains positive over the whole temperature range and reaches a decent value of $S = 80$ μV/K at $T = 300$ K. This is in sharp contrast to the negative Hall coefficient obtained on this sample [Fig. 3(b)], as well as previous ARPES band maps showing that $E_F$ lies above the Dirac point so that electron-type surface Dirac fermions dominate the transport [25]. With further reduction of $x$ to $x = 0.8$, the Seebeck coefficient shows a non-monotonic behavior with increasing temperature. It has a small negative value at low $T$ and changes sign at $T \sim 150$ K. The $x = 0.7$ sample exhibits a similar behavior, except that the negative $S$ is more pronounced and the crossover to positive value occurs at an even higher temperature.

The Seebeck and Hall coefficients of all the $(Bi_{1-x}Sb_x)_2Te_3$ samples with $0 \leq x \leq 1$ are summarized in Fig. 5 as a function of $T$ and $x$. The complex sign evolution dissects the phase diagram into three distinct regions: negative $S$ with negative $R_H$ (region I), positive $S$ with negative $R_H$ (region II), and positive $S$ with positive $R_H$ (region III). In most materials the sign of $S$ is consistent with the type of charge carriers estimated from the Hall effect [5,32,33]. In other words, if the sign of $S$ changes at a critical doping or temperature [34], the Hall coefficient is expected to reverse sign simultaneously. However, in the $(Bi_{1-x}Sb_x)_2Te_3$ TI films studied here, the sign consistency between the Seebeck coefficient and Hall coefficient is violated in certain region of the phase diagram.

## IV. THEORETICAL DISCUSSIONS

To understand the anomalous transport properties of the $(Bi_{1-x}Sb_x)_2Te_3$ films, we have performed first principle electronic structure calculations. Fig. 6 shows the calculated band structures of 5 QL $(Bi_{1-x}Sb_x)_2Te_3$ films along the K − Γ − M direction for $x = 0.75$, 0.90 and 1.0. The surface states can be clearly identified by two linear bands crossing at the Γ point. For $0.75 \leq x \leq 1.0$, the overall band structures are almost the same and the only differences happen at the relative positions between the Dirac point and the bulk VBM along the Γ − M direction. Therefore, it is reasonable to calculate the thermoelectric properties of various $(Bi_{1-x}Sb_x)_2Te_3$ samples by tuning the $E_F$ for different doping levels based on the band structure for $x = 0.90$.

The transport coefficients are calculated by the Landauer approach, in which the Seebeck coefficient $S$ is described as $S = -(k_B/e)(I_1/I_0)$. Here $k_B$ is the Boltzmann constant, $e$ is the elementary charge, and $I_n$ ($n = 0, 1$) is the dimensionless integrals given by $I_n = \int dx \frac{x^n e^x}{(e^x+1)^2} \overline{T}(x)$, in which $x = (E - E_F)/(k_B T)$ and $\overline{T}(x) = \overline{T}(E)$ is the electronic transmission function. $\overline{T}(E) = M(E) T(E)$. $M(E)$ is the distribution of conduction modes, which counts the number of conduction channels at a given energy $E$ and is proportional to the group velocity along the transport direction times the density of states. Fig. 7 shows the $M(E)$ for the 5 QL $(Bi_{1-x}Sb_x)_2Te_3$ film with $x = 0.9$ obtained by directly counting the conduction channels over the whole Brillouin zone. $T(E)$ is the transmission probability, which for diffusive transport is given by $T(E) = \lambda(E)/L$, where $\lambda(E)$ represents the mean free path and $L$ is the transport length. $T(E)$ is usually calculated using the approximation of constant scattering time or constant mean free path [35]. However, a dual scattering time model is necessary for TIs [17] because the topological surface states are protected from backscattering and thus have longer scattering time than the bulk states. Here we use $\lambda(E) = \lambda_1$ for surface states with energies within the bulk gap and $\lambda(E) = \lambda_2$ otherwise. The ratio $r = \lambda_1/\lambda_2$ is the only variable relevant to $S$, and should be larger than 1 to account for the topological protection of surface states. To reproduce the experimental $S$ for $x = 0.9$ at 300 K, $r = 5$ is selected. Notice that $r$ in reality is sample-dependent and might be different. However, according to our tests, this would not influence the calculated $S$ qualitatively, and thus would not affect our theoretical discussion that is kept at a qualitative level.

To understand the sign anomaly between the Seebeck and Hall coefficients, we calculate the Seebeck coefficient as a function of $T$ based on the band structure for $x = 0.9$ and vary the $E_F$ to mimic the influence of $x$ in experiment [23], as presented in Fig. 8. For $E_F$ below the bulk VBM (i.e., $E_F \leq 0$), $S$ is always positive and increases with increasing

$E_F$. In contrast, for $E_F > 0$ increasing $E_F$ leads to a downward shift and sign reversal of $S$ from positive to negative at low temperatures. The critical point $T_C$, where the sign change happens, gets higher for increasing $E_F$. All these features are qualitatively consistent with the experimental observations.

The underlying physical picture of the unexpected transport behavior of TIs can be illustrated intuitively by involving the contributions from both the surface and bulk states, as shown in Fig. 9(a). Specifically, the Seebeck coefficient can be expressed as:

$$S = (\sigma_s S_s + \sigma_b S_b)/(\sigma_s + \sigma_b). \tag{2}$$

Here $\sigma_{s,b}$ and $S_{s,b}$ represent the electrical conductivity and Seebeck coefficient of the surface and bulk states, respectively. We focus on the regime of $E_F$ above (and close to) the bulk VBM, where the interesting sign change of $S$ happens. In this regime $S_s$ is negative because the Dirac point is located below $E_F$, whereas $S_b$ is always positive due to the thermally activated hole-type bulk carriers. Their relative contribution to the total $S$ is controlled by the ratio $\sigma_b/\sigma_s$, which depends sensitively on $T$ and $E_F$. Specifically, decreasing $T$ and/or increasing $E_F$ (decreasing $x$ in experiments) would reduce $\sigma_b/\sigma_s$ because of the reduction of bulk carrier density. At a critical $T$ or $x$, the sign of $S$ will change from positive to negative, which is precisely the trend observed in experiment (see Fig. 5).

From Eq. (2) it is clear that the critical condition for changing the sign of $S$ is $\sigma_b/\sigma_s = |S_s/S_b|$ provided that $S_s$ and $S_b$ have opposite signs. What about the Hall coefficient $R_H$? In the presence of both surface and bulk conduction, $R_H$ can be derived from the two-band model as $R_H = (-\sigma_s \mu_s + \sigma_b \mu_b)/(\sigma_s + \sigma_b)^2$, where $\mu_{s,b}$ represents the carrier mobility.

The sign of $R_H$ is reversed when $\sigma_b/\sigma_s = \mu_s/\mu_b$. Interestingly, the critical conditions for changing the sign of $S$ and $R_H$ differ considerably in TI due to the unique properties of the Dirac-like SSs. The linear dispersion indicates weak electron-hole asymmetry around $E_F$, which leads to much smaller $S_s$ than $S_b$ so that $|S_s/S_b| \ll 1$ is easily satisfied. On the other hand, topological protection of the surface Dirac fermions strongly enhances its mobility [16], resulting in $\mu_s/\mu_b \gg 1$ as evidenced by previous experiment [36]. Therefore, there exists a sizeable region (region II in Fig. 5) in which the relationship $|S_s/S_b| < \sigma_b/\sigma_s < \mu_s/\mu_b$ is valid, causing the opposite sign of $S$ and $R_H$. In region I or III, the ratio $\sigma_b/\sigma_s$ is too small or too large such that either the surface or bulk states dominate both types of transports, giving the same sign for $S$ and $R_H$. Fig. 9(b) summarizes the theoretical phase diagram, which qualitatively reproduces the experimental one in Fig. 5 given the fact that in the experiments increasing $T$ and $x$ enhances $\sigma_b/\sigma_s$.

## V. CONCLUSION

To conclude, we observe a sign anomaly between the Seebeck and Hall coefficients in 5 QL $(Bi_{1-x}Sb_x)_2Te_3$ TI thin films over a range of Sb concentrations and temperatures. The origin is that *p*-type bulk states dominate the thermoelectric transport, whereas *n*-type SSs dominate the magnetoelectric transport. Our experiment establishes the importance of topological SSs in the thermoelectric effect of TI, which has been frequently emphasized in theoretical proposals [15-22]. Unlike the common two-band scenario, the surface and bulk channels of TI are separated in space, which enables the design of high-efficiency thermoelectric devices through novel heterojunction and superlattice architectures. The basic philosophy is to control the surface and bulk bands separately and take advantage of the optimal transport properties of each channel to further enhance *ZT* [20]. Moreover, the interactions between the surface and bulk states provides a new knob for improving *ZT*. It has been proposed recently that the surface-bulk coupling will lead to a strongly energy dependent lifetime for the SSs [17], which causes a maximum of $S_s$ and a significantly enhanced *ZT*. In this work our TI films are quite far away from the situation considered in the theory in Ref. 17. For future investigations we plan to tune the system into a regime dominated by the topological SSs to test these theoretical proposals.

Jinsong Zhang, Xiao Feng and Yong Xu contributed equally to this work. We acknowledge Wenhui Duan and Xiaobin Chen for suggestions and comments. This work was supported by the National Natural Science Foundation of China, the Ministry of Science and Technology of China and the Chinese Academy of Sciences.

**Figure captions**

FIG. 1 (color online). Schematic diagram for transport measurements and the results of 5 QL $Bi_2Te_3$ and $Sb_2Te_3$ films. (a) Schematic device for the thermoelectric measurements. (b) The sheet resistance ($R_\square$) of $Bi_2Te_3$ and $Sb_2Te_3$ shows metallic behavior at high $T$ and turns to weakly insulating at low $T$. (c) In $Bi_2Te_3$ and $Sb_2Te_3$ the $S_{xx}$ has a quasi-linear $T$ dependence, but with opposite signs. (d) The Hall coefficient of $Bi_2Te_3$ is negative, indicating electron-like charge carriers which contribute negative $S$; whereas in $Sb_2Te_3$ the hole-like charge carriers contribute positive $S$. The solid symbols are obtained by fitting the slope of field dependent Hall traces at fixed temperatures, and the lines are measured by switching the current and voltage contacts at fixed magnetic field when the temperature is slowly swept [31].

FIG. 2 (color online). The temperature dependence of the sheet resistance for the 5 QL $(Bi_{1-x}Sb_x)_2Te_3$ films with $0.7 \leq x \leq 0.97$. $R_\square$ shows an enhanced insulating tendency when $x$ is close to 0.95.

FIG. 3 (color online). The Hall effect for the 5 QL $(Bi_{1-x}Sb_x)_2Te_3$ films with $0.7 \leq x \leq 0.97$. (a) Field dependence of the Hall resistance measured at constant temperatures. (b) The Hall coefficient $R_H$ as a function of continuously increased (red lines) and constant (black squares) temperatures. For each sample the sign of $R_H$ is unchanged with increasing $T$ (negative for $x \leq 0.9$ and positive for $x \geq 0.95$).

FIG. 4 (color online). The Seebeck coefficient for the 5 QL $(Bi_{1-x}Sb_x)_2Te_3$ films with $0.7 \leq x \leq 0.97$. In samples with $x = 0.7$ and $0.8$, a sign reversal of $S$ from negative to positive is revealed as $T$ rises. As $x$ is further increased from $x = 0.9$ to $0.97$, the sign reversal totally disappears and a maximum positive value is found with $x = 0.95$. At high $T$, the sign anomaly between $S$ and $R_H$ is clearly revealed for $0.7 \leq x \leq 0.9$.

FIG. 5 (color online). The phase diagram of $(Bi_{1-x}Sb_x)_2Te_3$ films summarizing the $S_{xx}$ as a function of $x$ and $T$. The contours with $S = 0$ (wine dash line) and $R_H = 0$ (blue dash-dot line) divide this phase diagram into three regions: negative $S_{xx}$ and $R_H$ (region I), positive $S_{xx}$ but negative $R_H$ (region II), and positive $S$ and $R_H$ (region III).

FIG. 6 (color online). The band structure of 5 QL $(Bi_{1-x}Sb_x)_2Te_3$ calculated by the density functional theory with $x = 0.75$ (a), 0.9 (b), and 1.0 (c). Red dashed lines indicate the position of Dirac point, and black dashed lines represent the locations of the $\Gamma$ points. The bulk VBM is selected as the energy reference ($E = 0$).

FIG. 7 (color online). The calculated distribution of conduction modes $M(E)$ as a function of energy for the 5 QL $(Bi_{1-x}Sb_x)_2Te_3$ film with $x = 0.9$.

FIG. 8 (color online). The Seebeck coefficient as a function $T$ and $E_F$ calculated based on the band structure in Fig. 6(b). While $S_{xx}$ is always positive for $E_F \leq 0$, $S_{xx}$ changes from negative to positive with increasing $T$ for $E_F > 0$.

FIG. 9 (color online). (a) Schematic drawing shows the coexistence of electron-type surface Dirac fermions and hole-type bulk carriers in a TI thin film. (b) A theoretical phase diagram for TI thin films with n-type surface and p-type bulk charge carriers. When increasing $\sigma_b/\sigma_s$ (realized by increasing $T$ and Sb content in experiment), a negative-to-positive sign change happens for both $S$ and $R_H$ but at distinctly different conditions: $\sigma_b/\sigma_s = |S_s/S_b| \ll 1$ for $S$ and $\sigma_b/\sigma_s = \mu_s/\mu_b \gg 1$ for $R_H$. Three different regions (I/II/III) are defined according to the signs of $S$ and $R_H$, the same as in Fig. 5.

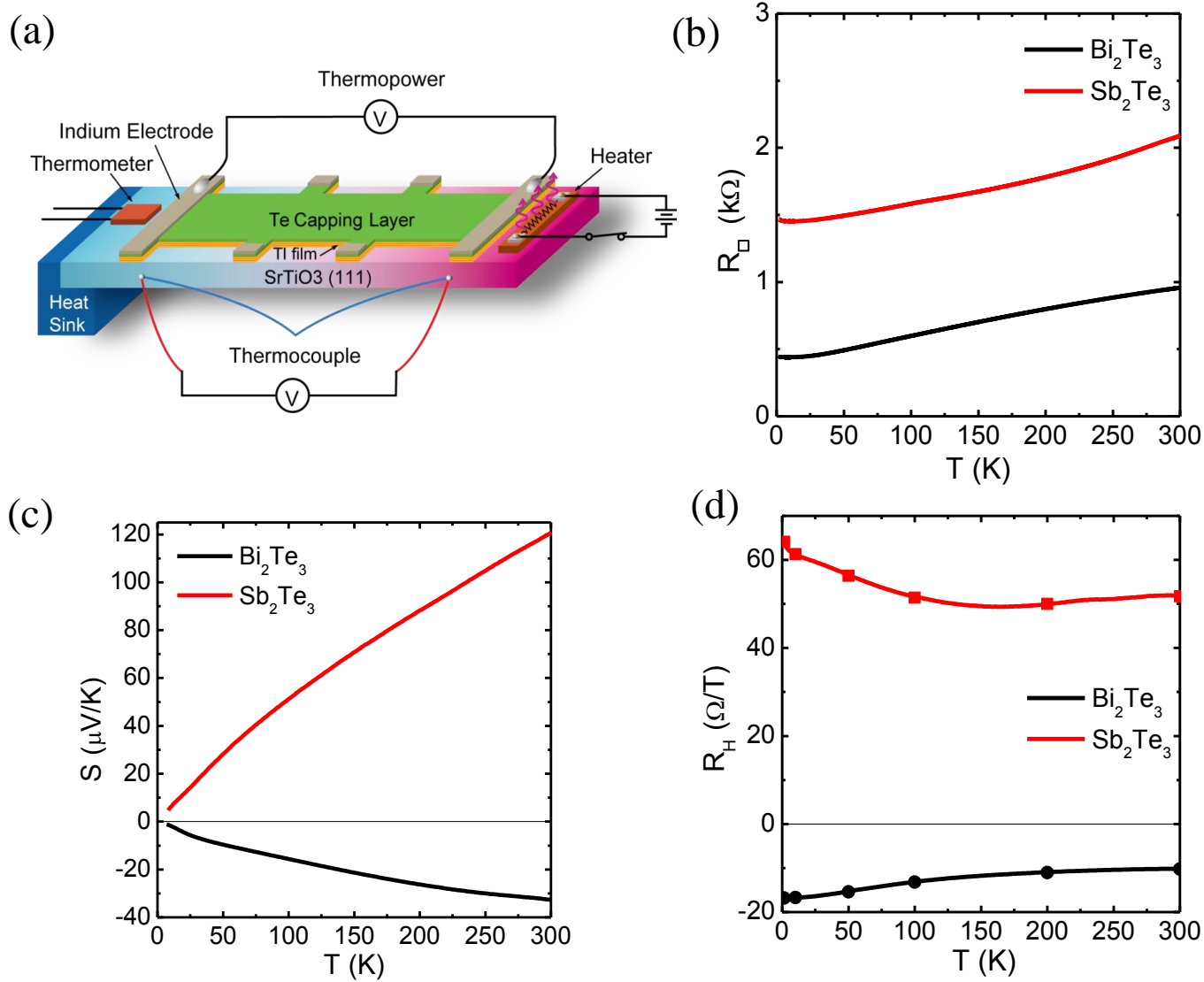

Figure 1

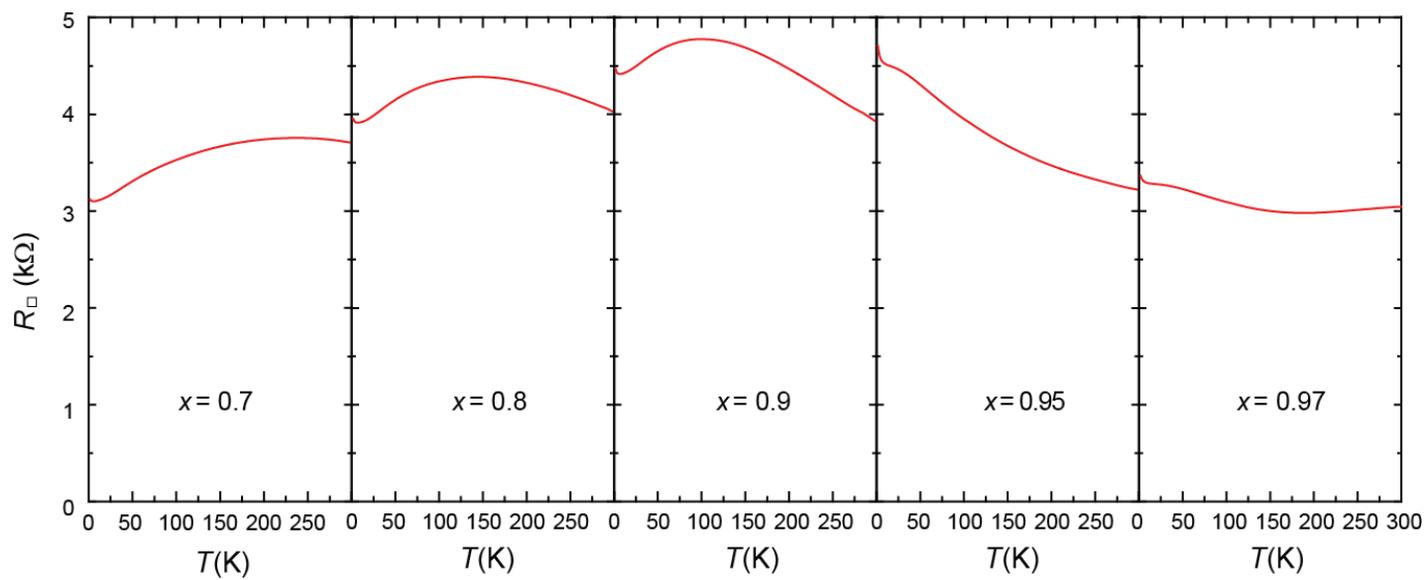

Figure 2

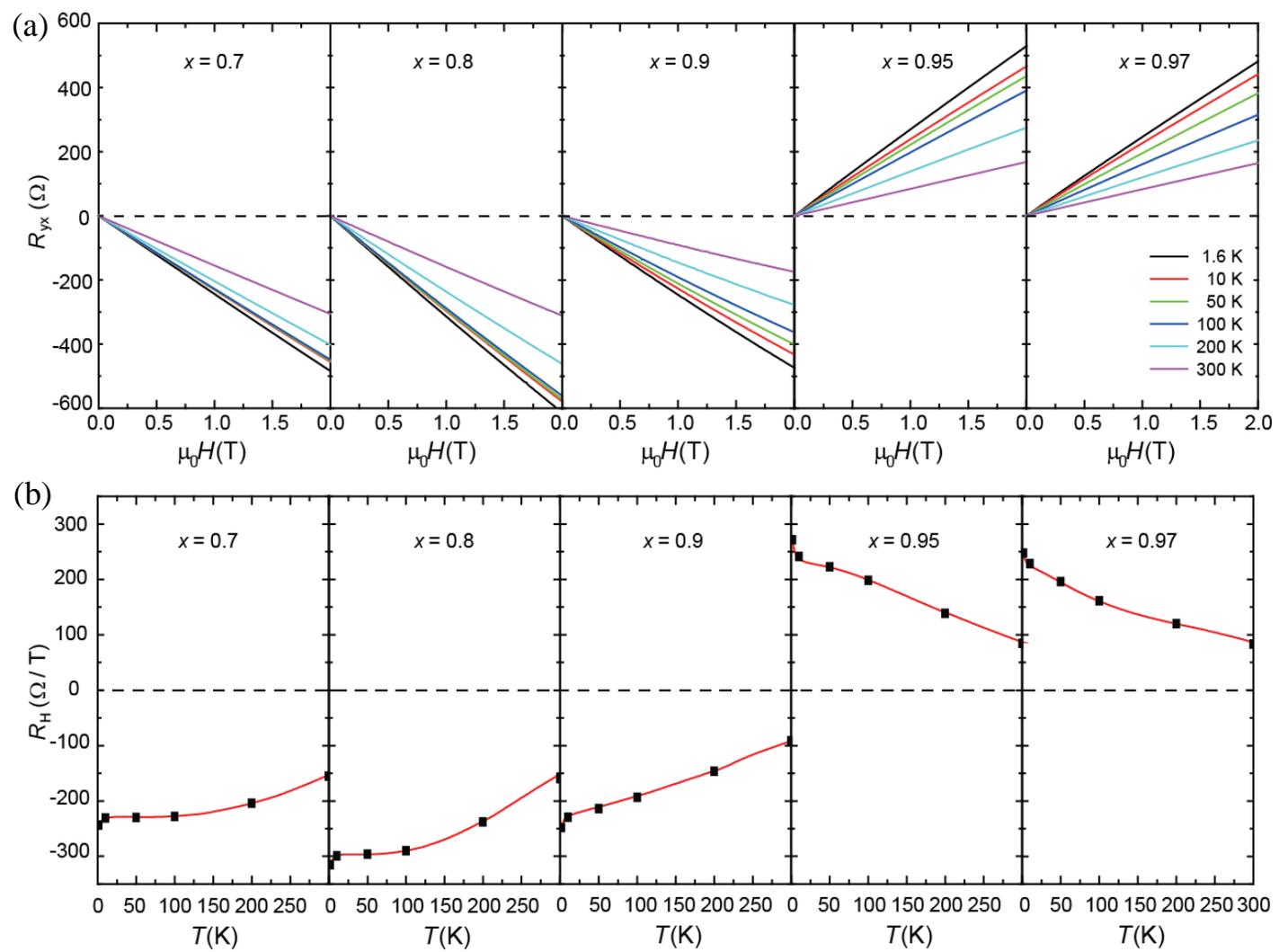

Figure 3

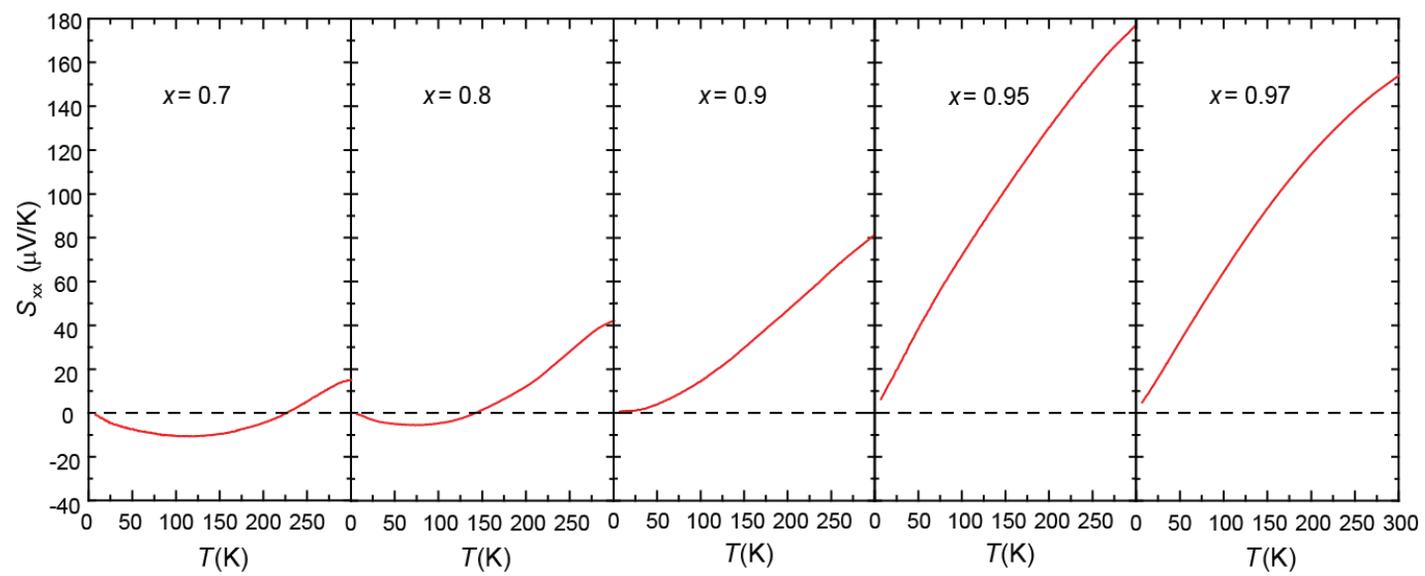

Figure 4

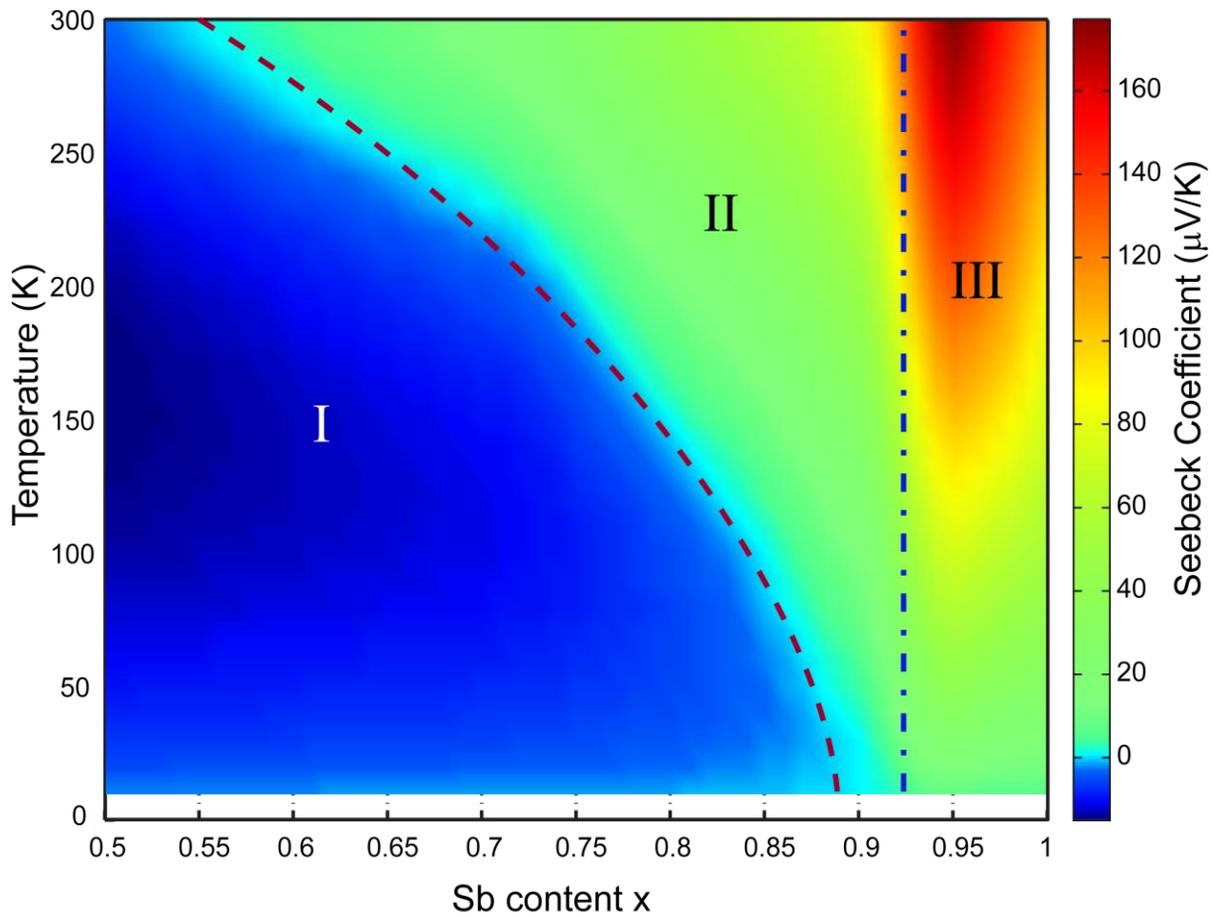

Figure 5

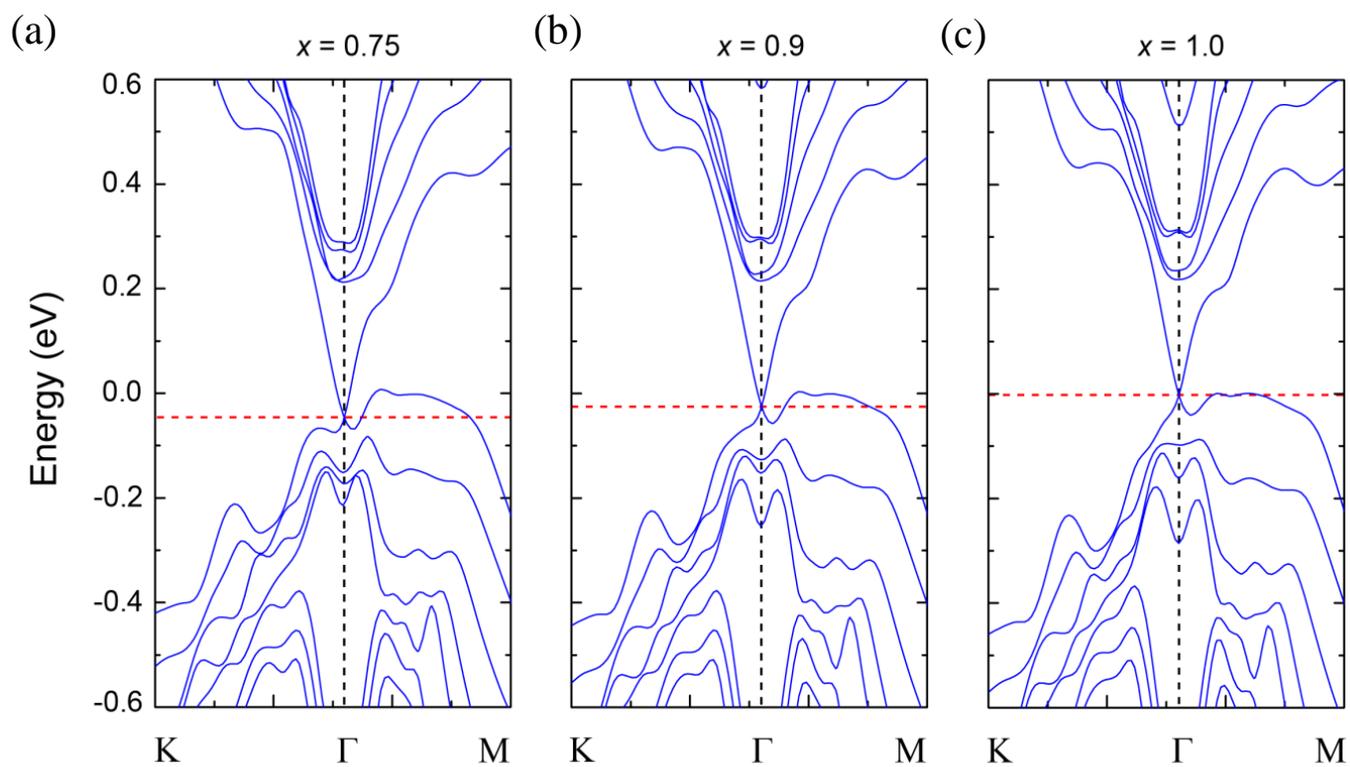

Figure 6

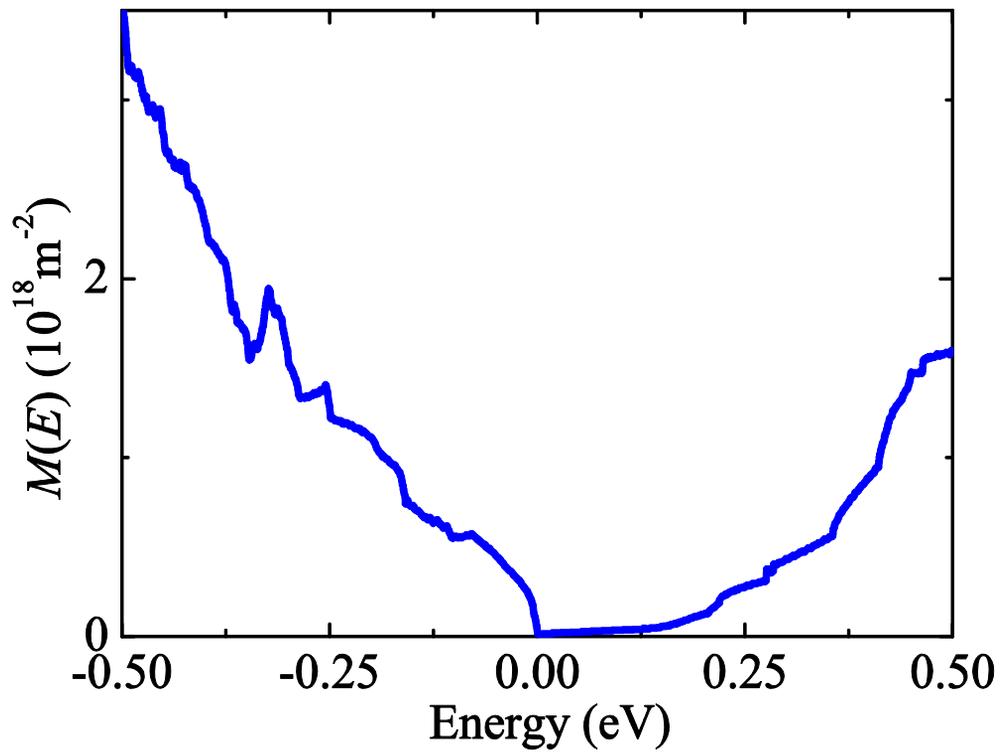

Figure 7

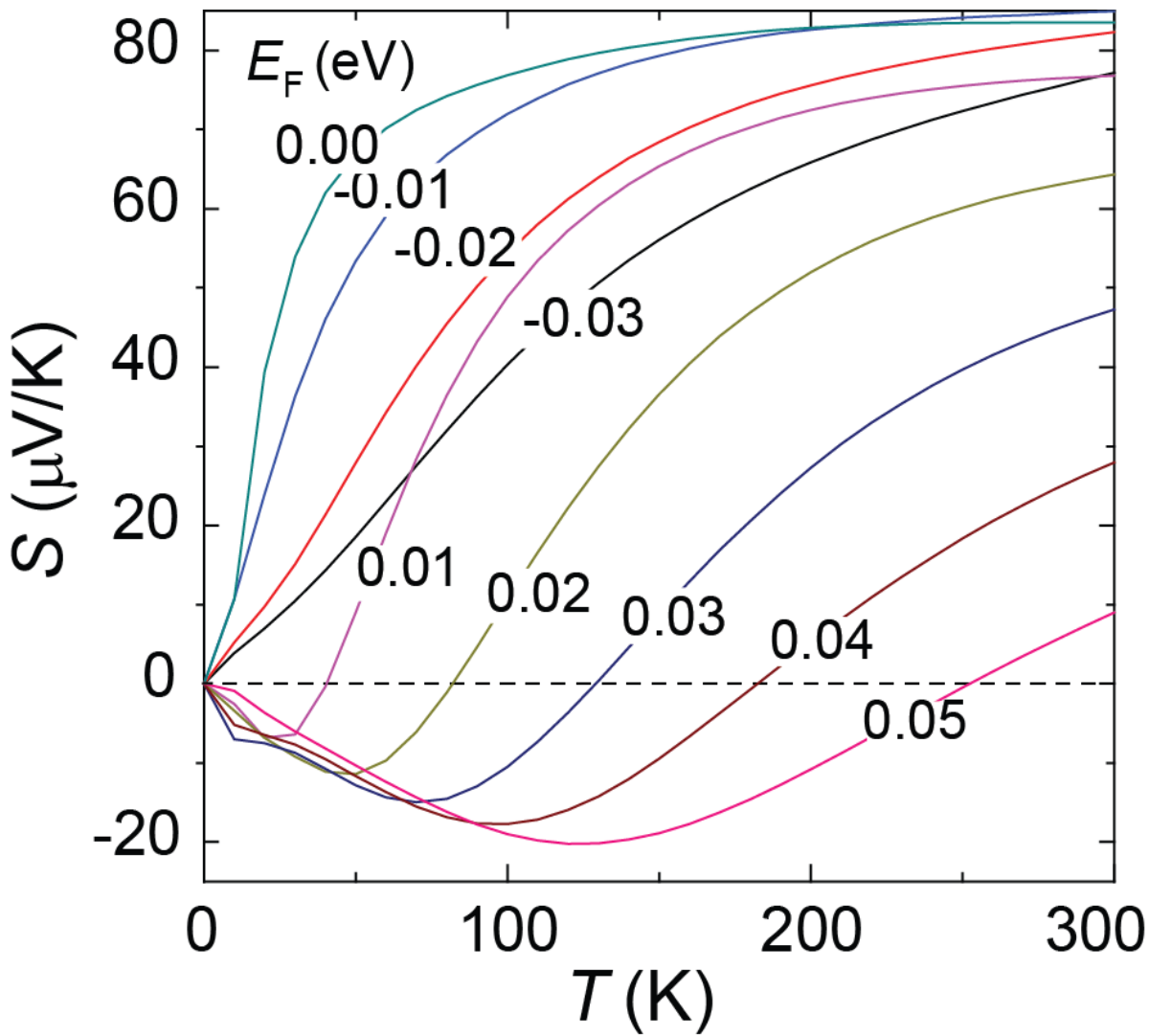

Figure 8

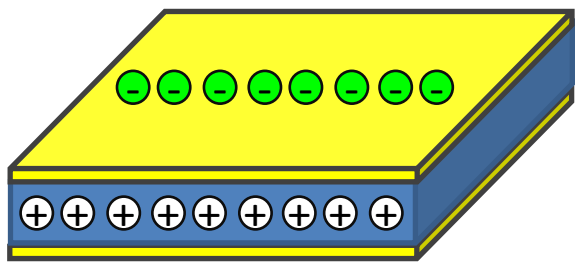 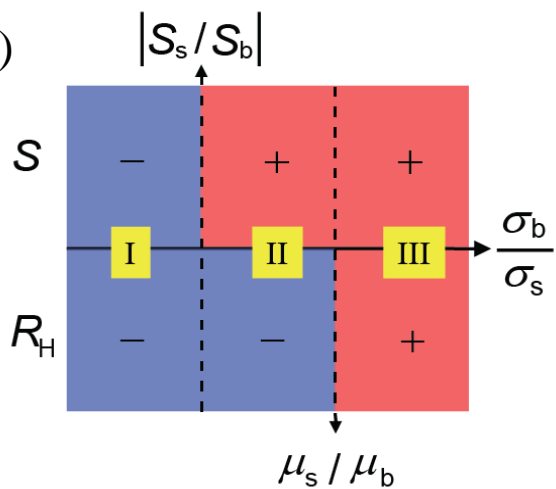

Figure 9